# Real-Time Power Electronics Control and Monitoring with TI F28379D DSC and GUI Composer

Ilyas Bennia, Lotfi Baghli, Ehsan Jamshidpour
lotfi.baghli@univ-lorraine.fr
GREEN, Université de Lorraine, Nancy, France
Abdelkader Mechernene
LAT, Université de Tlemcen, Tlemcen, Algeria

Jean-Philippe Martin
CNRS, LEMTA Université de Lorraine, Nancy, France
Driss Yousfi
ESETI, ENSAO Université Mohammed Premier, Oujda, Morocco

*Abstract*— **This paper details the implementation and experimental validation of a real-time control system for a three-phase induction motor using the Texas Instruments TMS320F28379D microcontroller. The system integrates pulse-width modulation (PWM) generation, analog-to-digital conversion (ADC), digital-to-analog conversion (DAC), and quadrature encoder feedback to facilitate precise control under various strategies. A current sensing solution based on the AMC1301 isolation amplifier and shunt resistor ensures accurate and safe current measurement for feedback loops. Two control algorithms, V/f and Field-Oriented Control (FOC) are implemented and tested. Real-time parameter tuning and data visualization are achieved using GUI Composer, enabling efficient system debugging and interaction. Experimental results demonstrate smooth speed reversal, fast dynamic response, and stable performance under both step and multi-step inputs. While GUI Composer effectively supports general monitoring and control, limitations in signal bandwidth are noted compared to professional-grade platforms. The results confirm the robustness and effectiveness of the implemented control strategies for high-performance induction motor applications.**

## I. INTRODUCTION

As power electronics increasingly shape the landscape of modern technology, from personal transportation, brushless DC motors used in e-bikes and two-wheel self-balancing scooters require precise speed and torque control[1]. In electric vehicles (EVs), traction inverters manage multi-phase induction or permanent magnet synchronous motors (PMSMs), while auxiliary systems rely on isolated DC/DC converters[2]. In aviation and drone technology, lightweight motor drives and compact inverters are essential for propulsion and stabilization. Renewable energy systems such as photovoltaic (PV) inverters employ power electronics for maximum power point tracking (MPPT), DC/AC conversion[3], and synchronization with the utility grid[4], [5]. Additionally, grid-forming inverters play a crucial role in microgrids operating in islanded mode[6], [7]. Each of these applications demands accurate real-time control, high-frequency PWM generation, fast ADC sampling, and robust communication protocols for seamless Human-Machine (HMI) and Machine-to-Machine (M2M) interaction[8]. As technology advances, there is also more pressure to reduce the overall cost of these systems[9].

In well-funded labs and industrial R&D centers, engineers typically turn to high-end rapid control prototyping (RCP) and hardware-in-the-loop (HIL) platforms like dSPACE MicroLabBox or OPAL-RT, which offer robust integration of control algorithms, real-time feedback, and high-level human–machine interfaces (HMI)[10]. However, these platforms come at a steep price, often exceeding €10,000–€20,000, rendering them unattainable for many universities, especially in emerging countries where budget constraints and limited access to specialized tools are common. This challenge is especially critical in academic and research environments, where engineers, students, and researchers must prototype, validate, and experiment within limited budgets. In this context, low-cost, flexible, and scalable real-time solutions emerge as ideal tools for driving innovation while also supporting hands-on learning and experimental development, helping to educate the next generation of embedded and control engineers[11].

Academic labs and students often experiment with Arduino or ESP32 boards to implement control for motors and converters. While these platforms are affordable and easy to program, they quickly reveal critical shortcomings in power electronics contexts: limited PWM resolution (typically 8–10 bits), unsynchronized and noisy ADCs, and the absence of real-time processing capabilities or floating-point computation. Such constraints make them unsuitable for tasks that require precise feedback loops, high-frequency switching, or synchronization with sensor data, such as encoder pulses or back-EMF signals.

Some developers have explored more capable 16-bit microcontrollers, such as the Microchip dsPIC33FJ128MC802, which is optimized for motor control and costs less than $5. Despite its efficiency, it lacks a hardware floating-point unit (FPU), limiting its performance in algorithms that rely on real-time trigonometric, PID, or field-oriented control (FOC) calculations[12], [13]. It also suffers from limited community support and lacks modular development environments or host boards tailored for power conversion experiments.

Similarly, the ESP32, a 32-bit dual-core processor with Wi-Fi and Bluetooth, has gained popularity for IoT and general embedded applications, but falls short in power electronics. It lacks ADC-to-PWM synchronization, has relatively poor analog accuracy, and cannot deliver the deterministic timing needed in motor drives, which demand low-latency interrupt handling and precise phase alignment.

This leaves a considerable gap between affordability and functionality—a gap that the TI C2000 Delfino F28379D aims to fill[14]. By offering high-resolution PWM modules, multiple synchronized ADCs, an encoder interface via eQEP, and a dual-core floating-point architecture at under $35, it presents a compelling case as a bridge between expensive prototyping systems and the limitations of entry-level microcontrollers. Moreover, its compatibility with C programming and the graphical user interface GUI Composer environment enables a low-cost SCADA-like experience that empowers students and researchers to design, monitor, and control real-time systems with professional-grade fidelity[15].

Although MATLAB Simulink includes a C2000 library that supports code generation for the F28379D[16], it falls short when it comes to advanced motor control tasks. Critical

features—such as real-time changes to PWM settings, adjusting action qualifiers on the fly, or deactivating specific inverter legs—are not supported in Simulink's auto-generated environment[17], [18]. These functions, however, are straightforward to implement in low-level C using direct register access. While several research works use Simulink with the F28379D for power electronics and algorithm validation through HIL, detailed, C-based implementations remain scarce.

In contrast to auto-generated code from environments like MATLAB Simulink, hand-written C code offers far more compact and efficient execution, thanks to the optimized instruction set of TI's Digital Signal Controllers (DSCs). Modern C development eliminates the need for low-level assembly—except in rare, time-critical routines—while maintaining full access to the controller's performance capabilities[19]. However, despite the F28379D's strong presence in academic and industrial projects[20], there remains a noticeable gap in openly available, register-level C code examples tailored for power electronics applications[21].

In this work, we aim to bridge that gap. By combining low-level C programming with Texas Instruments' GUI Composer, we propose a complete and cost-effective solution for real-time control and monitoring of power electronic systems. This approach provides not only high flexibility and performance but also serves as an accessible platform for researchers, educators, and students to experiment, learn, and develop real-world applications without relying on expensive prototyping tools.

## II. SYSTEM ARCHITECTURE

The system is structured as depicted in Fig.1 to serve as a versatile real-time control platform for various power electronic converters, including motor drives, DC/DC converters, and inverter systems. At its core is the F28379D controller capable of generating multiple high-resolution PWM signals required to drive switching devices such as IGBTs or MOSFETs. Each PWM signal is configurable in terms of frequency, dead-time insertion, and edge alignment, allowing support for different modulation strategies such as unipolar, bipolar, and space vector PWM. The feedback acquisition is handled through multiple synchronized ADC modules, which sample voltages and currents at precise time instants, typically triggered by PWM events to ensure minimal phase delay. This synchronization enables accurate implementation of closed-loop control algorithms such as current-mode control, field-oriented control (FOC), or voltage regulation. Encoder signals from rotating machines are captured using an integrated eQEP module, which provides position and speed information essential for vector control. The system also includes digital inputs for protection signals like overcurrent or fault flags, and digital outputs for hardware enable or interlock logic. Communication with a host PC is established via UART or USB, providing a channel for data logging, real-time parameter tuning, and monitoring through a SCADA-style interface. The control loop, implemented in C, runs periodically at a fixed sampling rate and performs tasks such as reading ADCs, computing control laws (e.g., PI controllers, Clarke-Park transforms), and updating PWM registers. This modular architecture makes the system adaptable for controlling different types of power converters with minimal hardware changes and ensures deterministic and low-latency operation required in high-performance power electronics control.

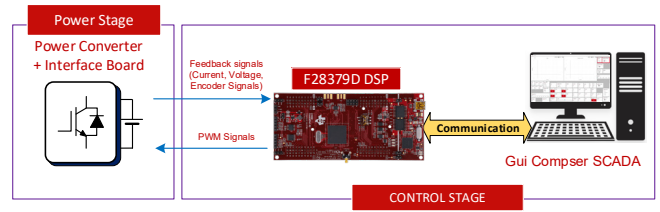

**Figure 1.** DSP-based architecture for real-time control of power converters

## III. TI F28379D INITIALIZATION & PERIPHERAL CONFIGURATION

The C code developed for the TMS320F28379D microcontroller was organized into a modular structure to ensure clarity, maintainability, and efficient debugging. The application was divided into multiple source and header files, each serving a specific role. The peripheral initialization routines and interrupt service routines (ISRs) were grouped in a dedicated implementation file. Meanwhile, global variables and control algorithms were placed in separate files to preserve modularity and ease of development.

The initialization and configuration of the microcontroller's peripherals were implemented in Code Composer Studio (CCS) using bit-field programming, which allowed for clear and direct manipulation of peripheral registers.

Initial setup included configuring the system clock and GPIOs, followed by the setup of the PWM and ADC modules to support precise and synchronized signal sampling. A dedicated interrupt-driven routine was implemented in which ADC conversions were triggered by PWM events, allowing analog signals such as current and voltage to be sampled at specific switching points. This synchronization improves data accuracy and ensures predictable behavior, essential in real-time control applications.

With this foundational structure, the code was made modular, readable, and well-suited for further development in embedded power electronic systems.

As can be seen in the following code, the system initialization is carried out in a modular and sequential manner within the main() function. Several initialization routines are declared beforehand, including functions for system control, GPIO setup, PWM generation, ADC and DAC configuration, and encoder interfacing. Each function is responsible for setting up a specific peripheral or subsystem. The system clock and peripheral clocks are first initialized, followed by GPIO configuration to assign the appropriate pin functions. The ePWM module is then configured to generate PWM signals, which also serve as triggers for the ADC. The ADC setup is split into two phases: general configuration and initialization, allowing for synchronized and interrupt-driven signal sampling. The DAC is initialized for monitoring purposes, and the quadrature encoder interface is prepared for position feedback. An informational debug message is printed upon successful startup. Finally, the program enters an infinite loop, ready for the main control logic to be executed, with all peripherals fully initialized and interrupt service routines defined for real-time responsiveness.

```c
#include "F28x_Project.h"
#include <math.h>
#include <stdio.h>
#include "control_alghorithme.h"
#include "variables.h"
// Declare global variables
// Function Prototypes
void initSystem();
void Gpio_setup();
void Init_PWM();
void Configure_ADC();
void Init_ADC();
void Configure_DAC();
void init_Encoder();
void InitSpi();

__interrupt void ADC_ISR(void);
// Main Function
void main(void)
{// Function call
 //Initialize System Control:
 // PLL, WatchDog, enable Peripheral Clocks
 initSystem();
 // Initialize GPIO
 // Set the direction and function of each GPIO pin
 Gpio_setup();
 // Initialize ePWM with a single comparator value
 Init_PWM();
// Initialize and configure the ADC and DAC
 Configure_ADC();
 Init_ADC();
 Configure_DAC();
 // Initialize and configure the EQEP
 init_Encoder();
 // Debug message
 printf("MSG_INFO: System started - Code Debugger Work Properly\n");

 while(1){
    // main control loop with control code
 }
}
```

The initSystem() function performs essential system-level initialization for the TMS320F28379D microcontroller. It starts by configuring the system control, including the PLL, watchdog timer, and enabling peripheral clocks via InitSysCtrl(). CPU interrupts are disabled globally to allow safe peripheral setup. The PIE (Peripheral Interrupt Expansion) control registers and vector table are then reset to default states. Next, the ADC interrupt service routine (ADC_ISR) is mapped to its corresponding PIE vector entry by temporarily enabling protected register access. The function then enables the ADC interrupt within the PIE interrupt enable register and activates the CPU interrupt group associated with PIE group 1. Finally, global and real-time interrupts are enabled, allowing the system to respond to interrupt events as soon as initialization completes.

```c
void initSystem(void)
{
    // Initialize system control: PLL, WatchDog, enable peripheral clocks
    InitSysCtrl();

    // Disable CPU interrupts
    DINT;

    // Initialize PIE control registers to their default state
    InitPieCtrl();
    // Disable CPU interrupts and clear all CPU interrupt flags
    IER = 0x0000;
    IFR = 0x0000;

    // Initialize the PIE vector table with default ISR pointers
    InitPieVectTable();

    // Map ISR functions
    EALLOW; // Allow protected register access
    PieVectTable.ADCA1_INT = &ADC_ISR;    // ADC interrupt (Group 1, INTx1)
    EDIS;   // Disable protected register access
    // Enable PIE interrupts for ADC and Timer0 (Group 1)
    PieCtrlRegs.PIEIER1.bit.INTx1 = 1; // Enable ADCA1_INT
    // Enable CPU interrupt for PIE Group 1
    IER |= M_INT1; // Enable CPU interrupt for Group 1
    // Enable global and real-time interrupts
    EINT;   // Enable global interrupts (INTM)
    ERTM;   // Enable real-time interrupt (DBGM)
}
```

The Gpio_setup() function configures the GPIO pins of the TMS320F28379D microcontroller for their specific peripheral functions. Pull-up resistors are disabled on PWM-related pins GPIO0 to GPIO5, and these pins are multiplexed to serve as ePWM outputs (EPWM1A/B, EPWM2A/B, EPWM3A/B). Additionally, GPIO20 to GPIO22 are configured for the eQEP1 encoder interface to enable quadrature position feedback signals. GPIO32 is set as a general-purpose output pin, with its pull-up resistor disabled, and it is cleared initially for debugging purposes. The function uses protected register access to safely modify GPIO control registers and concludes with a debug message confirming successful GPIO setup.

```c
void Gpio_setup()
{
    EALLOW;
// Enable PWM1-3 on GPIO0-GPIO5,
    GpioCtrlRegs.GPAPUD.bit.GPIO0 = 1;    // Disable pull-up on GPIO0 (EPWM1A)
    GpioCtrlRegs.GPAPUD.bit.GPIO1 = 1;    // Disable pull-up on GPIO1 (EPWM1B)
    GpioCtrlRegs.GPAMUX1.bit.GPIO0 = 1;    // Configure GPIO0 as EPWM1A
    GpioCtrlRegs.GPAMUX1.bit.GPIO1 = 1;    // Configure GPIO1 as EPWM1B
    GpioCtrlRegs.GPAPUD.bit.GPIO2 = 1;    // Disable pull-up on GPIO2 (EPWM2A)
    GpioCtrlRegs.GPAPUD.bit.GPIO3 = 1;    // Disable pull-up on GPIO3 (EPWM2B)
    GpioCtrlRegs.GPAMUX1.bit.GPIO2 = 1;    // Configure GPIO2 as EPWM2A
    GpioCtrlRegs.GPAMUX1.bit.GPIO3 = 1;    // Configure GPIO3 as EPWM2B
    GpioCtrlRegs.GPAPUD.bit.GPIO4 = 1;    // Disable pull-up on GPIO4 (EPWM3A)
    GpioCtrlRegs.GPAPUD.bit.GPIO5 = 1;    // Disable pull-up on GPIO5 (EPWM3B)
    GpioCtrlRegs.GPAMUX1.bit.GPIO4 = 1;    // Configure GPIO4 as EPWM3A
    GpioCtrlRegs.GPAMUX1.bit.GPIO5 = 1;    // Configure GPIO5 as EPWM3B
    // Enable GPIO pins for eQEP1 encoder
    GpioCtrlRegs.GPAMUX2.bit.GPIO20 = 1;   // GPIO20 -> eQEP1A
    GpioCtrlRegs.GPAMUX2.bit.GPIO21 = 1;   // GPIO21 -> eQEP1B
    GpioCtrlRegs.GPAMUX2.bit.GPIO22 = 1;   // GPIO22 -> eQEP1 Index
    // GPIO 32 for debugging purposes
    GpioCtrlRegs.GPBPUD.bit.GPIO32 = 1;    // Disable pull-up on GPIO32
    GpioCtrlRegs.GPBMUX1.bit.GPIO32 = 0;   // Configure GPIO32 as GPIO
    GpioCtrlRegs.GPBDIR.bit.GPIO32 = 1;    // GPIO32 = output
    GpioDataRegs.GPBCLEAR.bit.GPIO32 = 1;  // Clear

    printf("MSG_INFO: code started - Gpio_setup Properly\n");
}
```

The Init_PWM() function initializes and configures three PWM modules (EPwm1, EPwm2, and EPwm3) on the

TMS320F28379D microcontroller, which is commonly used in motor control and power electronics applications.

- **Duty Cycle Initialization:**
  The function starts by setting the duty cycles of all three PWM channels to 0.5 (50%). This means that the PWM outputs will have equal ON and OFF times, producing a neutral or "safe torque" state where no net torque or flux rotation occurs, preventing abrupt changes or damage at startup.
- **Stopping PWM Clocks:**
  It disables the time-base clocks (TBCLKSYNC = 0) to stop all PWM timers while configuring the modules, preventing unintended or unstable outputs during setup.
- **Time-Base Configuration:**
  For each PWM module (EPwm1, EPwm2, EPwm3), the following settings are applied:
  -The period register (TBPRD) is set to PwmTBPRD, which defines the PWM frequency.
  -The phase register (TBPHS) is set to zero, meaning no phase shift is applied.
  -The counter (TBCTR) is reset to zero.
  -The counter mode is set to up-down counting (TB_COUNT_UPDOWN), which produces a symmetrical triangular waveform, useful for balanced PWM signals.
    -Phase synchronization is disabled (PHSEN = TB_DISABLE) since each PWM operates independently.
    -Clock prescalers (HSPCLKDIV and CLKDIV) are set to divide by 1, meaning the PWM clock runs at full peripheral clock speed.
- **Compare Submodule Configuration:**
  The compare registers (CMPA and CMPB) determine the duty cycle of each PWM output. They are set based on the initialized duty cycles (50%), scaled by the period value (PwmTBPRD). Both Compare A and B are set equal to maintain symmetrical behavior on the complementary outputs.
- **Shadowing Configuration:**
  Shadow registers are enabled (CC_SHADOW) for both compare registers to ensure duty cycle updates occur synchronously with the PWM timer, avoiding glitches or mid-cycle changes. The load mode is set to load the shadow values only when the timer reaches zero (CC_CTR_ZERO), providing stable and predictable PWM updates.
- **Action Qualifier Configuration:**
  The Action Qualifier determines how the PWM outputs behave during the timer up and down counts:
  -For PWM A (EPWMxA): The output is set high on the up count compare event (CAU = AQ_SET) and cleared low on the down count compare event (CAD = AQ_CLEAR).
  -For PWM B (EPWMxB): The output is cleared low on up count compare (CBU = AQ_CLEAR) and set high on down count compare (CBD = AQ_SET). This produces complementary PWM signals on A and B outputs, essential for driving half-bridge circuits.
- **Dead-Band Generation:**
  To protect switching devices (like MOSFETs or IGBTs) from shoot-through (both switches on the same leg conducting simultaneously), a dead-band generator is enabled with full output mode (DB_FULL_ENABLE). The polarity is set for active high complementary outputs (DB_ACTV_HIC), and the dead-time rising (DBRED) and falling (DBFED) edges are both set to a predefined delay (PWMdeadBand). This introduces a small delay between turning one transistor off and the complementary transistor on, preventing short circuits.
- **Forcing PWM Outputs Low:**
  Before starting the PWM normally, the outputs (PWM A and B) for each module are forced low (AQCSFRC.bit.CSFA = 1 and CSFB = 1). This ensures the outputs are in a safe, known state until the controller is ready to enable them.
- **Re-enabling PWM Clocks:**
  After all PWM modules are fully configured, the time-base clocks are re-enabled (TBCLKSYNC = 1), starting the timers and thus the PWM signals simultaneously. This synchronization ensures coherent PWM outputs across the three modules, crucial in multi-phase motor control or multi-leg inverters.

In summary, this function carefully configures three PWM channels with complementary outputs, dead-band protection, synchronized timing, and safe initial conditions to be ready for motor or power electronics control.

```c
void Init_PWM()
{
    // Equal duty cycles means no rotation of flux
    DutyCycles[0]=0.5;//Safty Torque Operation
    DutyCycles[1]=0.5;//Safty Torque Operation
    DutyCycles[2]=0.5;//Safty Torque Operation
// Parameterize the Configuration
    EALLOW;
    ClkCfgRegs.PERCLKDIVSEL.bit.EPWMCLKDIV = 0;   // Set ePWM Clock Divider (0 = /1, 1 = /2)
    CpuSysRegs.PCLKCR0.bit.TBCLKSYNC = 0;         // Stop all ePWM time-base clocks
    EDIS;
    // Initialize Time-Base Submodule
    EPwm1Regs.TBPRD = PwmTBPRD;                   // Set the time-base period TBPRD
    EPwm1Regs.TBPHS.bit.TBPHS = 0x0000;           // Set phase to 0
    EPwm1Regs.TBCTR = 0x0000;                     // Clear counter
    EPwm1Regs.TBCTL.bit.CTRMODE = TB_COUNT_UPDOWN; // Set count mode to up-down
    EPwm1Regs.TBCTL.bit.PHSEN = TB_DISABLE;       // Disable phase synchronization
    EPwm1Regs.TBCTL.bit.HSPCLKDIV = TB_DIV1;      // High-speed clock pre-scaler
    EPwm1Regs.TBCTL.bit.CLKDIV = TB_DIV1;         // Clock pre-scaler
    // Configure Compare Submodule
    EPwm1Regs.CMPA.bit.CMPA = DutyCycles[0]* PwmTBPRD; // Set Compare A value
    EPwm1Regs.CMPB.bit.CMPB = DutyCycles[0]* PwmTBPRD; // Set Compare B value (same as CMPA)
    // Configure Shadowing
    EPwm1Regs.CMPCTL.bit.SHDWAMODE = CC_SHADOW;   // Enable shadow mode for Compare A
    EPwm1Regs.CMPCTL.bit.SHDWBMODE = CC_SHADOW;   // Enable shadow mode for Compare B
    EPwm1Regs.CMPCTL.bit.LOADAMODE = CC_CTR_ZERO; // Load Compare A on Zero
    EPwm1Regs.CMPCTL.bit.LOADBMODE = CC_CTR_ZERO; // Load Compare B on Zero
    // Configure Action Qualifier Submodule
    EPwm1Regs.AQCTLA.bit.CAU = AQ_SET;            // Set PWM output on up count
    EPwm1Regs.AQCTLA.bit.CAD = AQ_CLEAR;          // Clear PWM output on down count
    EPwm1Regs.AQCTLB.bit.CBU = AQ_CLEAR;          // Clear PWM B on up count
    EPwm1Regs.AQCTLB.bit.CBD = AQ_SET;            // Set PWM B on down count
    // Enable Dead-Band Generator
    EPwm1Regs.DBCTL.bit.OUT_MODE = DB_FULL_ENABLE; // Enable dead-band generation
    EPwm1Regs.DBCTL.bit.POLSEL = DB_ACTV_HIC;     // Configure active high complementary
    EPwm1Regs.DBRED.bit.DBRED = PWMdeadBand;      // Set dead-band rising edge delay
    EPwm1Regs.DBFED.bit.DBFED = PWMdeadBand;      // Set dead-band falling edge delay
```

```c
    // Turn Off PWM waiting for the flaginhibitPwm to activate it
    EPwm1Regs.AQCSFRC.bit.CSFA = 1; // Force PWM1A low
    EPwm1Regs.AQCSFRC.bit.CSFB = 1; // Force PWM1B low
    // Initialize Time-Base Submodule
    EPwm2Regs.TBPRD = PwmTBPRD;                    // Set the time-base period TBPRD
    EPwm2Regs.TBPHS.bit.TBPHS = 0x0000;            // Set phase to 0
    EPwm2Regs.TBCTR = 0x0000;                      // Clear counter
    EPwm2Regs.TBCTL.bit.CTRMODE = TB_COUNT_UPDOWN; // Set count mode to up-down
    EPwm2Regs.TBCTL.bit.PHSEN = TB_DISABLE;        // Disable phase synchronization
    EPwm2Regs.TBCTL.bit.HSPCLKDIV = TB_DIV1;       // High-speed clock pre-scaler
    EPwm2Regs.TBCTL.bit.CLKDIV = TB_DIV1;          // Clock pre-scaler

    // Configure Compare Submodule
    EPwm2Regs.CMPA.bit.CMPA = DutyCycles[1]* PwmTBPRD; // Set Compare A value
    EPwm2Regs.CMPB.bit.CMPB = DutyCycles[1]* PwmTBPRD; // Set Compare B value (same as CMPA)

    // Configure Shadowing
    EPwm2Regs.CMPCTL.bit.SHDWAMODE = CC_SHADOW;    // Enable shadow mode for Compare A
    EPwm2Regs.CMPCTL.bit.SHDWBMODE = CC_SHADOW;    // Enable shadow mode for Compare B
    EPwm2Regs.CMPCTL.bit.LOADAMODE = CC_CTR_ZERO;  // Load Compare A on Zero
    EPwm2Regs.CMPCTL.bit.LOADBMODE = CC_CTR_ZERO;  // Load Compare B on Zero

    // Configure Action Qualifier Submodule
    EPwm2Regs.AQCTLA.bit.CAU = AQ_SET;             // Set PWM output on up count
    EPwm2Regs.AQCTLA.bit.CAD = AQ_CLEAR;           // Clear PWM output on down count
    EPwm2Regs.AQCTLB.bit.CBU = AQ_CLEAR;           // Clear PWM B on up count
    EPwm2Regs.AQCTLB.bit.CBD = AQ_SET;             // Set PWM B on down count

    // Enable Dead-Band Generator
    EPwm2Regs.DBCTL.bit.OUT_MODE = DB_FULL_ENABLE; // Enable dead-band generation
    EPwm2Regs.DBCTL.bit.POLSEL = DB_ACTV_HIC;      // Configure active high complementary
    EPwm2Regs.DBRED.bit.DBRED = PWMdeadBand;       // Set dead-band rising edge delay
    EPwm2Regs.DBFED.bit.DBFED = PWMdeadBand;       // Set dead-band falling edge delay

    EPwm2Regs.AQCSFRC.bit.CSFA = 1; // Force PWM2A low
    EPwm2Regs.AQCSFRC.bit.CSFB = 1; // Force PWM2B low

    // Initialize Time-Base Submodule
    EPwm3Regs.TBPRD = PwmTBPRD;                    // Set the time-base period TBPRD
    EPwm3Regs.TBPHS.bit.TBPHS = 0x0000;            // Set phase to 0
    EPwm3Regs.TBCTR = 0x0000;                      // Clear counter
    EPwm3Regs.TBCTL.bit.CTRMODE = TB_COUNT_UPDOWN; // Set count mode to up-down
    EPwm3Regs.TBCTL.bit.PHSEN = TB_DISABLE;        // Disable phase synchronization
    EPwm3Regs.TBCTL.bit.HSPCLKDIV = TB_DIV1;       // High-speed clock pre-scaler
    EPwm3Regs.TBCTL.bit.CLKDIV = TB_DIV1;          // Clock pre-scaler

    // Configure Compare Submodule
    EPwm3Regs.CMPA.bit.CMPA = DutyCycles[2]* PwmTBPRD; // Set Compare A value
    EPwm3Regs.CMPB.bit.CMPB = DutyCycles[2]* PwmTBPRD; // Set Compare B value (same as CMPA)

    // Configure Shadowing
    EPwm3Regs.CMPCTL.bit.SHDWAMODE = CC_SHADOW;    // Enable shadow mode for Compare A
    EPwm3Regs.CMPCTL.bit.SHDWBMODE = CC_SHADOW;    // Enable shadow mode for Compare B
    EPwm3Regs.CMPCTL.bit.LOADAMODE = CC_CTR_ZERO;  // Load Compare A on Zero
    EPwm3Regs.CMPCTL.bit.LOADBMODE = CC_CTR_ZERO;  // Load Compare B on Zero

    // Configure Action Qualifier Submodule
    EPwm3Regs.AQCTLA.bit.CAU = AQ_SET;             // Set PWM output on up count
    EPwm3Regs.AQCTLA.bit.CAD = AQ_CLEAR;           // Clear PWM output on down count
    EPwm3Regs.AQCTLB.bit.CBU = AQ_CLEAR;           // Clear PWM B on up count
    EPwm3Regs.AQCTLB.bit.CBD = AQ_SET;             // Set PWM B on down count

    // Enable Dead-Band Generator
    EPwm3Regs.DBCTL.bit.OUT_MODE = DB_FULL_ENABLE; // Enable dead-band generation
    EPwm3Regs.DBCTL.bit.POLSEL = DB_ACTV_HIC;      // Configure active high complementary
    EPwm3Regs.DBRED.bit.DBRED = PWMdeadBand;       // Set dead-band rising edge delay
    EPwm3Regs.DBFED.bit.DBFED = PWMdeadBand;       // Set dead-band falling edge delay

    EPwm3Regs.AQCSFRC.bit.CSFA = 1; // Force PWM3A low
    EPwm3Regs.AQCSFRC.bit.CSFB = 1; // Force PWM3B low

    // Synchronize Time-Base Clocks
    EALLOW;
    CpuSysRegs.PCLKCR0.bit.TBCLKSYNC = 1;          // Enable all ePWM time-base clocks
    EDIS;
}
```

The Configure_ADC() function sets up the core operating parameters of the ADC modules (ADCA and ADCB) on the TMS320F28379D. First, it configures the ADC clock by setting the prescaler (PRESCALE = 6), which divides the system clock by 4 to generate a suitable ADCCLK. Then, both ADC modules are powered up by setting the ADCPWDNZ bit to 1, followed by a 1 ms delay (DELAY_US(1000)) to allow the analog circuits to stabilize. The resolution is configured to 12-bit (RESOLUTION = 0) with a differential input mode (SIGNALMODE = 1), which is sufficient for general analog measurements. Additionally, the interrupt pulse position is set to late (INTPULSEPOS = 1), meaning that any ADC interrupt will be triggered after the conversion result is available, ensuring accurate and complete data reads.

The Init_ADC() function handles the setup of the sampling process and triggering mechanism. It begins by defining an acquisition window (ACQPS = 30), which corresponds to a 75 ns sample time—adequate for 12-bit resolution. The analog input channels are selected: ADCA is configured to sample from channel A2, and ADCB from channel B2, each with the same acquisition window and triggered by ePWM1's SOCA signal (TRIGSEL = 5). The ADC is set to generate an interrupt (INT1E = 1) when SOC0 completes (INT1SEL = 0). On the PWM side, ePWM1 is configured to issue a start-of-conversion trigger on every period event (SOCAEN = 1, SOCASEL = ET_CTR_PRD, and SOCAPRD = ET_1ST), thereby synchronizing ADC sampling with the PWM signal for time-aligned data acquisition.

```c
void Configure_ADC()
{
  EALLOW;
  //write configurations
```

```c
    AdcaRegs.ADCCTL2.bit.PRESCALE   =  6;    //set ADCCLK divider to /4
    AdcbRegs.ADCCTL2.bit.PRESCALE   =  6;    //set ADCCLK divider to /4
    // Power up the ADC
    AdcaRegs.ADCCTL1.bit.ADCPWDNZ = 1;
    AdcbRegs.ADCCTL1.bit.ADCPWDNZ = 1;
    DELAY_US(1000);  // Wait 1ms for ADC stabilization
    // Set resolution and signal mode
    AdcaRegs.ADCCTL2.bit.RESOLUTION   =  0;    // 12-bit resolution var.h type must be unsigned 12, not 16
    AdcaRegs.ADCCTL2.bit.SIGNALMODE = 1;  // Differential mode
    AdcbRegs.ADCCTL2.bit.RESOLUTION   =  0;    // 12-bit resolution
    AdcbRegs.ADCCTL2.bit.SIGNALMODE = 1;
    //Set pulse positions to late
    AdcaRegs.ADCCTL1.bit.INTPULSEPOS = 1;
    AdcbRegs.ADCCTL1.bit.INTPULSEPOS = 1;
         // 0: Early Pulse The interrupt is generated at the start of the conversion process.
         // 1: The interrupt is generated at the EOC , after the result is available in the result register
    EDIS;
    printf("MSG_INFO: code started - ADC configured Properly\n");
}

void Init_ADC()

{
Uint16 acqps;
    // Determine minimum acquisition window (in SYSCLKS) based on resolution
    acqps = 30; //75ns   for ADC_RESOLUTION_12BIT

    //Select the channels to convert and the end of conversion flag
    EALLOW;
    AdcaRegs.ADCSOC0CTL.bit.CHSEL = 2;   // A2-A3 pair
    AdcaRegs.ADCSOC0CTL.bit.ACQPS =   acqps;   //sample window is 100 SYSCLK cycles
    AdcaRegs.ADCSOC0CTL.bit.TRIGSEL = 5;  //trigger on ePWM1 SOCA/C

// ADCINB0,B1 VDC, IDC
    AdcbRegs.ADCSOC0CTL.bit.CHSEL = 2;   // B2-B3 pair
    AdcbRegs.ADCSOC0CTL.bit.ACQPS = acqps; //sample window is 100 SYSCLK cycles
    AdcbRegs.ADCSOC0CTL.bit.TRIGSEL = 5;  //trigger on ePWM1 SOCA/C
//Intrruption init

    AdcaRegs.ADCINTSEL1N2.bit.INT1SEL = 0; //end of SOC0 will set INT1 flag "source of the interruption"
    AdcaRegs.ADCINTSEL1N2.bit.INT1E = 1;    //enable INT1 flag

    // ePWM1 SOC at each period
    EPwm1Regs.ETSEL.bit.SOCAEN   = 1;              // Enable SOC on A group
    EPwm1Regs.ETSEL.bit.SOCASEL  = ET_CTR_PRD;  // SOC on time-base counter equal to zero (TBCTR =TBPRD)
    EPwm1Regs.ETPS.bit.SOCAPRD   = ET_1ST;            // Generate pulse on 1st event
    EDIS;
    printf("MSG_INFO: code started - ADC initialized Properly\n");

}
```

The Configure_DAC function initializes the DACB module of the TMS320F28379D to output analog voltages. First, it enables write access to protected registers using EALLOW. It then sets the DAC reference voltage to VREFHI (DACREFSEL = 1), ensuring the DAC output will scale between 0 and VREFHI (typically 3V). The load mode is set to immediate using the system clock (LOADMODE = 0), so any new DAC value written to DACVALS is reflected on the output immediately. The output buffer is powered on by enabling DACOUTEN, and the initial output value is set to 0 (DACVALS = 0), ensuring the DAC starts with a known state. A short delay (DELAY_US(10)) is added to allow time for the DAC circuitry to stabilize before use. Finally, access to protected registers is disabled with EDIS, and a debug message confirms successful configuration.

```c
void Configure_DAC()

{
    EALLOW;
    DacbRegs.DACCTL.bit.DACREFSEL = 1;// Use VREFHI as reference
    DacbRegs.DACCTL.bit.LOADMODE = 0; // Load on SYSCLK
    DacbRegs.DACOUTEN.bit.DACOUTEN = 1;// Power up the buffered DAC
    DacbRegs.DACVALS.all = 0;
    DELAY_US(10); // Delay for buffered DAC to power up
    EDIS;
    printf("MSG_INFO: code started - DAC configured Properly\n");

}
```

The init_Encoder() function initializes the eQEP1 (Enhanced Quadrature Encoder Pulse) module for reading position and speed from a quadrature encoder. The configuration starts by enabling access to protected registers using EALLOW. The encoder is set to operate in quadrature count mode (QSRC = 0), with extended count resolution enabled (XCR = 1), allowing detection of all edges of the encoder signals (4x resolution). The eQEP module is activated (QPEN = 1), and the unit time period is defined by setting QUPRD to 2,000,000 clock cycles, which corresponds to a 10 ms period at a 200 MHz system clock—used for regular sampling of position and speed. The initial position is reset (QPOSINIT = 0), and the maximum count is set to 4095, which suits a 1024-slot encoder using 4x decoding (4 × 1024 – 1 = 4095). For position management, the position counter is reset on the rising edge of the index pulse (QEPI) using PCRM = 0 and IEI = 2. The position latch is enabled on unit time-out events (QCLM = 1), and these events are also enabled (UTE = 1) to allow regular updates for velocity calculations. Finally, access to protected registers is disabled with EDIS, and a confirmation message is printed.

```c
void init_Encoder()

{
    EALLOW;
    // Configure the Quadrature Encoder (QEP) module
    EQep1Regs.QDECCTL.bit.QSRC = 0;     // Quadrature count mode
    EQep1Regs.QEPCTL.bit.QPEN = 1;      // Enable eQEP module for quadrature decoding
    EQep1Regs.QDECCTL.bit.XCR = 1;    // Enable extended count resolution (quadrature decoding)
    // Set Unit Period
    EQep1Regs.QUPRD= 2000000; // Set QUPRD register to 2000000 for 100 Hz (10 ms period) sysclk 200MHZ

    // Initialize position and set maximum position count
    EQep1Regs.QPOSINIT = 0;                // Initialize position counter to 0
    EQep1Regs.QPOSMAX = 4095;         // Set max position 4 * ENCODER_SLOTS-1 (4095 for 1024 encoder)
    //EQep1Regs.QPOSMAX =  0xffffffff;

    // Configure position counter reset and latching
    EQep1Regs.QEPCTL.bit.IEI = 2;      // Initializes the position counter on the rising edge of the QEPI signal //(QPOSCNT = QPOSINIT)
    EQep1Regs.QEPCTL.bit.PCRM = 0;    // Reset position counter on rising edge of QEPI signal
```

```
    EQep1Regs.QEPCTL.bit.QCLM = 1;         // Capture position latch on unit time-out event
    EQep1Regs.QEPCTL.bit.UTE = 1;          // Enable unit time-out event for periodic position capture
    EDIS;
    printf("MSG_INFO:code started - Encoder initialized Properly\n");
}
```

The ADC_ISR() function is an interrupt service routine triggered by the ADC conversion complete event. Upon entering, it enables global interrupts (EINT) to allow nested interrupts if needed. It sets GPIO pin 32 high (GPBSET.bit.GPIO32 = 1) as a debugging signal to indicate the ISR is running. The ADC interrupt flag for ADC interrupt 1 is cleared (ADCINTFLGCLR.bit.ADCINT1 = 1) to acknowledge the interrupt and prevent immediate retriggering. The routine then executes control functions based on the current value of the countCurrent variable: if countCurrent is 0, it calls Current_Control(), and if countCurrent is 1, it calls Speed_Control(). After running the appropriate control, it increments countCurrent, resetting it to zero when it reaches DS_IReg, effectively cycling through control stages. The ISR finally acknowledges the interrupt at the PIE controller level (PIEACK_GROUP1) to enable further interrupts from group 1, and clears GPIO pin 32 (GPBCLEAR.bit.GPIO32 = 1) to mark the ISR's end.

```
__interrupt void ADC_ISR()
{
   EINT;
   GpioDataRegs.GPBSET.bit.GPIO32 = 1;   // Set
   // Clear INT flag for this PWM
   AdcaRegs.ADCINTFLGCLR.bit.ADCINT1 = 1; // Clear ADC Interrupt 1 flag
   // Acknowledge this interrupt to receive more interrupts from group 1
switch(countCurrent)
 {
   case 0 :   Current_Control();
     break;
   case 1 :   Speed_Control();
     break;
 }
 if (++countCurrent==DS_IReg)  countCurrent=0;

PieCtrlRegs.PIEACK.all = PIEACK_GROUP1;
GpioDataRegs.GPBCLEAR.bit.GPIO32 = 1;   // Clear}
```

Finally, to ensure proper operation of the current and speed control algorithms in a power electronics system, the following commands must be implemented within the control loop. These commands serve as the critical interface between the control algorithm and the physical system, providing real-time feedback and updating actuation signals:

```
// PWM update
EPwm1Regs.CMPA.bit.CMPA = DutyCycles[0]* PwmTBPRD;
EPwm2Regs.CMPA.bit.CMPA = DutyCycles[1]* PwmTBPRD;
EPwm3Regs.CMPA.bit.CMPA = DutyCycles[2]* PwmTBPRD;
```

These lines update the PWM compare registers with new duty cycles calculated by the control algorithm. This step translates control decisions (e.g., voltage references or torque commands) into actual switching signals for the inverter.
```
Ias = (AdcaResultRegs.ADCRESULT0-OffsetIa)*GainI;
Ibs = (AdcbResultRegs.ADCRESULT0-OffsetIb)*GainI;
```
Here, the measured phase currents are read from the ADC result registers, corrected with offset and gain. These current measurements are essential inputs for the current control algorithm, allowing it to compare actual currents with reference values and regulate them accordingly.
This line reads the current position from the quadrature encoder. The position feedback is used in speed and rotor angle estimation, which is required for control algorithms such as Field-Oriented Control (FOC). This allows proper alignment of current components and accurate torque production.
```
// retrieve the position
newCount = EQep1Regs.QPOSCNT;
```
In summary, these commands establish the real-time feedback and actuation link necessary for closed-loop control. Without them, the control algorithm cannot respond to changes in system behavior or apply corrections, making them essential for both current and speed control in modern power electronics systems.

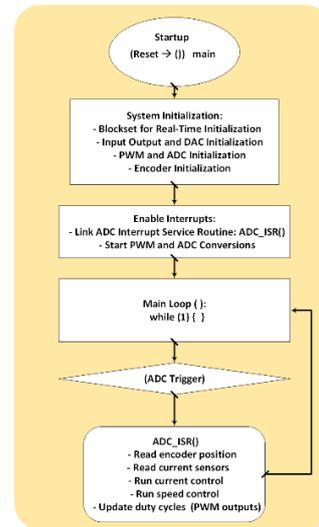

**Figure 2.** DSP- Real-time control execution flow using C code

In summary, the main program structure consists of several key components: a main function, an Interrupt Service Routine (ISR), and multiple initialization and configuration functions as illustrated in Fig.2. The main function serves as the entry point of the program, where it sequentially calls the configuration functions responsible for setting up peripherals such as the ADC, DAC, encoder, and PWM modules. It also installs and enables the necessary interrupts to handle time-critical events. Once all initializations are complete, the main function enters an infinite loop, maintaining the program's execution and waiting for interrupts to occur. The ISR function is triggered at precise intervals corresponding to the sampling period, ensuring timely processing of sensor data and execution of control algorithms such as current and speed control. This interrupt-driven architecture enables the program to respond efficiently and predictably to hardware events, maintaining real-time control of power electronics devices.

## IV. APPLICATION FOR MOTOR CONTROL

To test the implemented control code for the three-phase induction motor, the setup shown in Fig. 3 was constructed. The program was initially developed and compiled using Code Composer Studio, ensuring proper functionality and optimization. After building the code, it is flashed onto the target microcontroller using the GUI Composer, which serves as an interactive graphical interface for real-time control and monitoring. GUI Composer, provided by Texas Instruments, allows users to create custom dashboards that visually represent variables such as motor speed, current, torque, and other key parameters. This tool facilitates seamless communication with the embedded system via protocols like USB or Ethernet, enabling users to send commands, adjust control setpoints, and observe the dynamic response of the motor in real-time. The visual feedback helps in debugging, tuning control loops, and verifying system behavior under various operating conditions without the need for recompiling or redeploying the firmware. Its drag-and-drop interface and scripting capabilities simplify the creation of complex control panels tailored to specific applications, making it an invaluable resource for iterative development and experimental validation of motor control algorithms.

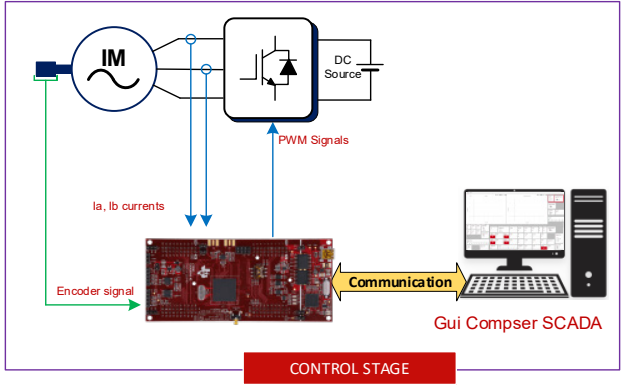

**Figure 3**. TI F28379d D DSP-motor control configuration

To sense current using the TI F28379D, which has an ADC input range of 0 to 3 V, the AMC1301 isolation amplifier is used in conjunction with a shunt resistor as exposed in Fig.4. The shunt resistor generates a small differential voltage (typically ±250 mV) proportional to the current flowing through the power circuit. This voltage is fed into the high side of the AMC1301, which provides galvanic isolation between the high-voltage power stage and the low-voltage control stage. The AMC1301 outputs a differential voltage centered around 1.25 V, well within the F28379D's ADC range, ensuring safe and accurate current measurement. This setup not only isolates the control circuitry from potentially dangerous high voltages but also conditions the signal appropriately for the ADC to process.

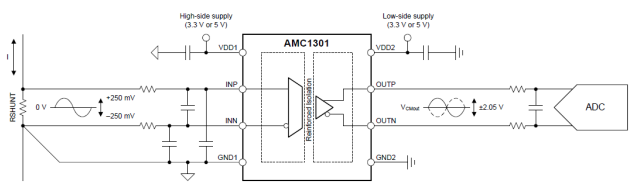

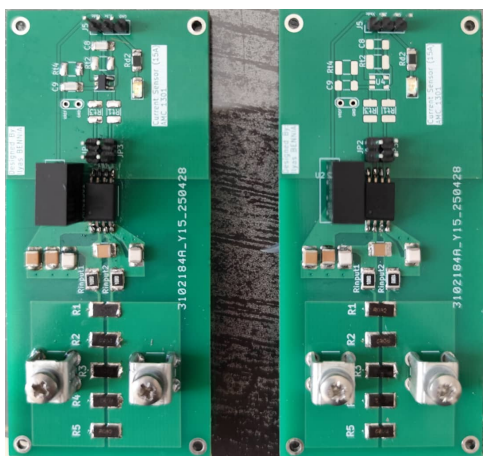

**Figure 4.** Current sensors schematic and prototype-based AMC 1301

During the experimental phase, multiple tests were conducted using both V/f control and Field-Oriented Control (FOC) strategies to evaluate the performance of the implemented control algorithms. These tests involved varying the motor speed across a wide range of setpoints to observe the dynamic and steady-state responses under different load and command conditions. Additionally, speed reversal scenarios were carried out to assess the system's stability and ability to handle direction changes smoothly without introducing significant torque ripples or instability. By analyzing the behavior of the motor during acceleration, deceleration, and direction inversion, we were able to validate the robustness and effectiveness of each control method.

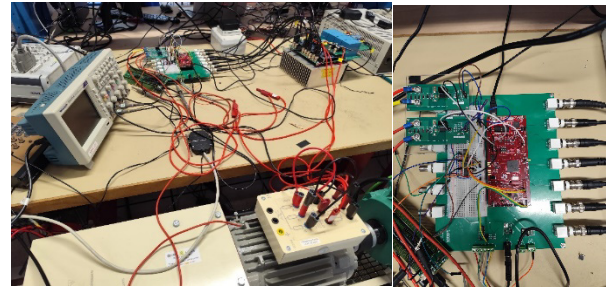

**Figure 4**. TI F28379d D DSP-motor control experimental setup

GUI Composer offers several advantages when used for real-time monitoring and control of embedded systems, particularly in motor control applications. Its intuitive drag-and-drop interface allows for the rapid development of custom dashboards that display key variables such as speed, current, and torque in a visually appealing and user-friendly format. The tool simplifies communication with the microcontroller through USB or Ethernet, enabling real-time parameter tuning and command execution without the need to recompile or redeploy the firmware. This greatly accelerates the development cycle and facilitates iterative testing. However, during our experiments, we encountered some limitations—most notably, the restricted bandwidth for signal visualization. Unlike professional platforms such as dSPACE ControlDesk, which offer high-speed data acquisition and oscilloscope-like signal tracing, GUI Composer struggles to accurately capture fast signal transitions. This limitation makes it difficult to observe high-frequency dynamics or transient responses in real time, which are crucial for tasks such as tuning current or speed control loops. As a result, while GUI Composer is effective for general-purpose monitoring and basic control, it is less suited for applications that require high-resolution timing analysis or detailed signal behavior during rapid events.

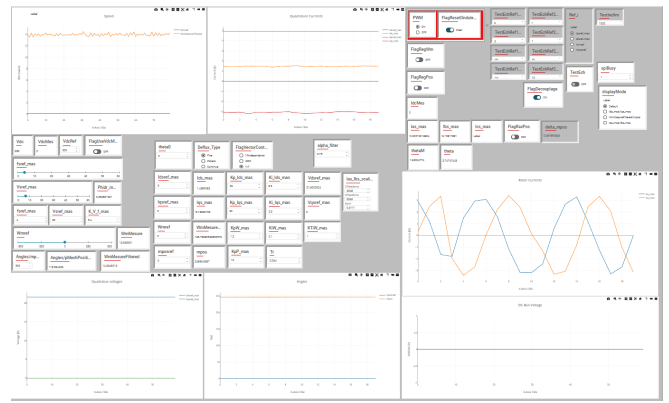

**Figure 5**. GUI composer developed a control and monitoring dashboard

Figure 5 shows the GUI Composer control dashboard used for real-time interaction with the motor control system. It comprises multiple interactive components, including live plots for monitoring variables such as speed, phase currents, and voltages, as well as numeric displays for key parameters like rotor angles. The dashboard also features control elements such as sliders, switches, and buttons that allow users to start or stop the motor, switch between control strategies (e.g., V/f and FOC), and adjust reference values or controller gains on the fly without the need to recompile the code. This real-time interaction significantly enhances the development and testing process by enabling quick tuning and immediate visual feedback.

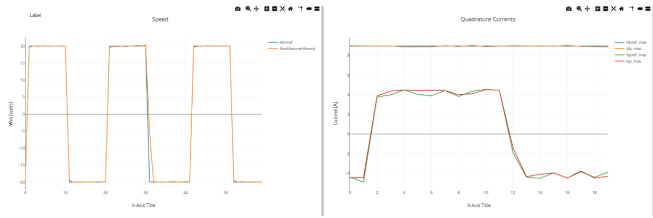

**Figure 6**. Speed, direct, and quadrature curves during the speed reverse test under FOC

Fig.6 illustrate the results of a speed inversion test for a three-phase induction motor under vector control FOC. The left graph shows the motor speed response, while the right graph presents the corresponding direct and quadrature (Id-Iq) currents over time.

In the speed plot, the reference speed is reversed periodically, switching between positive and negative setpoints. The actual speed closely follows the reference, indicating effective tracking and good dynamic response from the control system. The system handles speed inversion smoothly with minimal overshoot or delay, confirming that the control algorithm is well-tuned and responsive.

In the quadrature current plot, the Iq current rises as the speed reference increases, reflecting the torque demand needed to accelerate the motor. When the speed is reversed, the Iq current also inverts its polarity accordingly, demonstrating proper handling of torque direction changes. The current shows some ripple during transitions, which is typical during dynamic changes, but overall remains stable and controlled throughout the test. This analysis confirms that the control system successfully manages direction changes, maintaining system stability and accurate torque production, which are critical indicators of a robust FOC implementation.

Fig.7 illustrates the speed step response of a motor controlled using Field-Oriented Control (FOC). The reference speed undergoes a step change from 0 to 120 rad/s, and the measured speed effectively tracks this reference. The system demonstrates a fast dynamic response, reaching steady-state within approximately 0.5 seconds, with negligible overshoot and minimal steady-state error, indicating that the speed controller is well-tuned. The measured speed is filtered to reduce noise, which introduces a slight phase lag at the beginning of the response but does not significantly affect the tracking performance. Overall, the FOC system shows excellent tracking capability, smooth behavior, and stable performance during the speed step.

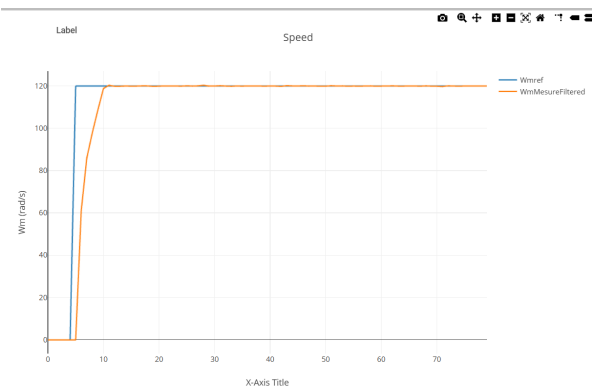

**Figure 7**. Step response under FOC

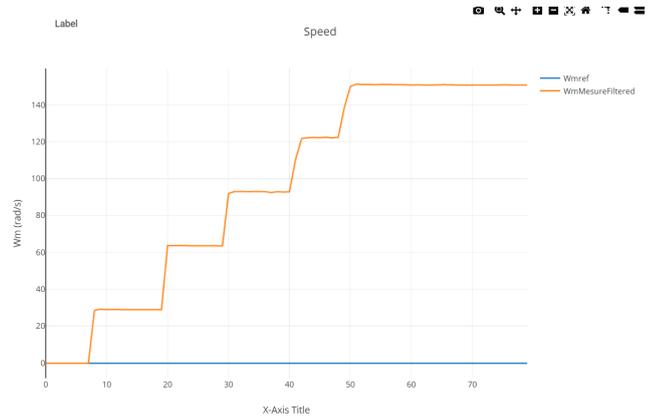

**Figure 8**. Step response under FOC

Fig.8 presents the response of a Field-Oriented Control (FOC) system to a multi-step speed reference input. In this case, the measured speed clearly shows multiple step increases, indicating that the system was subjected to successive speed commands. The motor responds to each step with a fast and smooth transient, quickly reaching the new steady-state speed with minimal overshoot and steady-state error after each transition. The filtering applied to the measured speed effectively smooths out noise while maintaining good responsiveness. Overall, the motor exhibits stable, fast, and well-controlled speed tracking under successive step changes, demonstrating the effectiveness of the FOC controller, although the reference signal display should be checked for proper plotting.

## V. CONCLUSIONS AND PERSPECTIVES

In this work, a complete real-time motor control system for a three-phase induction motor was developed and validated using the TMS320F28379D platform. The system incorporates essential peripherals, including PWM, ADC, DAC, and encoder modules, along with isolated current sensing using the AMC1301. Control algorithms for both scalar (V/f) and vector (FOC) strategies were implemented and evaluated under dynamic operating conditions. GUI Composer provided a user-friendly interface for real-time interaction, allowing on-the-fly tuning and system monitoring.

Experimental tests confirmed the controller's ability to achieve accurate speed tracking, smooth direction reversal, and stable torque production. FOC, in particular, exhibited excellent dynamic performance and robustness. However, the limited signal acquisition bandwidth of GUI Composer highlighted the need for higher-speed alternatives when analyzing fast transients or tuning inner control loops.

Overall, the implemented platform demonstrates a reliable and flexible solution for research and industrial applications in motor control. Future work may focus on integrating high-bandwidth data acquisition tools and expanding the system for multi-axis or sensorless control configurations.

### ACKNOWELDGMENT

The MG-FARM project has received funding from the European Union's Horizon 2020 Research and Program under Grant Agreement 963530.